\documentclass[11pt,twoside]{article}

\usepackage{asp2006}
\usepackage{epsf}
\usepackage{psfig}
\usepackage{lscape}

\begin{document}

\title{Early Type Galaxies in the Mid Infrared}
\author{A. Bressan\altaffilmark{1,2,3},
P. Panuzzo\altaffilmark{1},
L. Silva\altaffilmark{4},
L. Buson\altaffilmark{1},
M. Clemens\altaffilmark{1},
G.L. Granato\altaffilmark{1,2},
R. Rampazzo\altaffilmark{1},
J.R. Valdes\altaffilmark{3},
O. Vega\altaffilmark{3}}
\affil{$^{1}$ INAF Osservatorio Astronomico di Padova, vicolo dell'Osservatorio 5, 35122 Padova, Italy}
\affil{$^{2}$ SISSA, via Beirut 4, 34014, Trieste, Italy}
\affil{$^{3}$ INAOE, Luis Enrique Erro 1, 72840, Tonantzintla, Puebla, Mexico}
\affil{$^{4}$ INAF Osservatorio Astronomico di Trieste, Via Tiepolo 11, I-34131 Trieste, Italy}

\begin{abstract} 
We are performing a systematic study of the properties of early-type galaxies in the mid infrared
spectral region with the {\it Spitzer Space Telescope}. We present here high S/N {\it Spitzer} IRS
spectra of 17 Virgo early-type galaxies. Thirteen objects of the sample (76\%) show a pronounced
broad feature (above 10$\mu$m) which is spatially extended and likely of stellar origin. We argue
that this feature is (mostly) due to silicate emission from circumstellar envelopes of asymptotic
giant branch (AGB) stars. The remaining 4 objects, namely NGC 4486, NGC 4636, NGC 4550 and NGC
4435, are characterized by different levels and type of activity.
\end{abstract}




\section{Introduction}
With the advent of the {\it Spitzer Space Telescope} new frontiers have been opened in the study of
the stellar population content of early-type galaxies. In fact by means of mid infrared (MIR)
observations it is possible to estimate the age and metallicity of the old stellar populations and
to quantify both the presence of intermediate age stellar populations as well as even tiny amounts
of ongoing star formation.

The MIR spectral region of old and intermediate age stellar populations is affected by the presence
of mass-losing AGB giants. In fact, the dusty envelopes of oxygen-rich AGB stars show a clear
emission feature around 10$\mu$m and Bressan, Granato \& Silva (1998) and  Piovan, Tantalo \&
Chiosi (2003), suggested that this emission feature should be clearly seen in relatively old
populations. The studies by Bressan et al. (1998, 2001) have also shown that the MIR spectral
region could be used to bypass the effects of the age-metallicity  degeneracy because as the
observed system gets younger and/or its metallicity increases the feature initially gets larger and
then it dilutes into a  more broad emission that extends up to 15$\mu$m. Contrary to optical
indicators (e.g. colors or Balmer absorption lines), the above MIR feature shows an opposite
behavior with respect to age and metallicity variations. As a consequence it can be used in
combination with optical and UV observations to remove the age-metallicity degeneracy and to obtain
an accurate ranking of the stellar population ages.

Ongoing star formation is easily
detected in the MIR, from the presence of prominent
emission features such as PAHs and atomic or molecular
emission lines (e.g. Kaneda et al. 2005, Bressan et al. 2006a,b,
Panuzzo et al. 2007).
Moreover MIR nebular lines are a strong diagnostic
to disentangle star formation and AGN activity and
they also allow a {\sl direct and perhaps unique} determination of
the chemical abundance of the surrounding gas (Panuzzo et al. 2007).

On this basis we started a systematic study of the properties of early-type galaxies in the mid
infrared spectral region with the {\it Spitzer Space Telescope}. Here we report on the results
obtained with {\it Spitzer} IRS\footnote{The IRS was a collaborative venture between Cornell
University and Ball Aerospace Corporation, funded by NASA through the Jet Propulsion Laboratory and
the Ames Research Center}(Houck et al. 2004, Werner et al. 2004) MIR spectroscopic observations of
a sample of early-type galaxies in the Virgo cluster (Bressan et al 2006a,b).

\begin{table}
\centering
\caption{Virgo galaxies observed with IRS}
\label{tab1}
\scriptsize
\begin{tabular}{lcccclcccc}
\hline
\hline
Name & V$_T$ & SL1/2 & LL2    &S/N & Name & V$_T$ & SL1/2  & LL2 & S/N \\
     &       &cycles & cycles &    &      &       & cycles & cycles\\
\hline
NGC~4339  & 11.40  & 20 &  14  &  39  & NGC~4486  &  8.62  &  3 &   3  &  80     \\
NGC~4365  &  9.62  &  3 &  3   &  57  & NGC~4550  & 11.50  & 20 &  14  &  42     \\
NGC~4371  & 10.79  &  9 &  10  &  40  & NGC~4551  & 11.86  & 20 &  14  &  47    \\
NGC~4377  & 11.88  & 12 &   8  &  54  & NGC~4564  & 11.12  &  4 &   6  &  51     \\
NGC~4382  &  9.09  &  3 &   3  &  59  & NGC~4570  & 10.90  &  3 &   5  &  42     \\
NGC~4435  & 10.66  &  3 &   5  &  35  & NGC~4621  &  9.81  &  3 &   3  &  63     \\
NGC~4442  & 10.30  &  3 &   3  &  46  & NGC~4636  &  9.49  &  3 &   5  &  30     \\
NGC~4473  & 10.06  &  3 &   3  &  55  & NGC~4660  & 11.11  &  3 &   5  &  40     \\
NGC~4474  & 11.50  & 20 &  14  &  38  &           &        &         \\
\hline
\end{tabular}
\end{table}


\section{Observations and data reduction}
Standard staring mode short (SL1 and SL2) and long (LL2) low resolution IRS spectral observations
of 18 early-type galaxies, were obtained between January and July 2005. The galaxies were selected
among those that define the color-magnitude relation of the Virgo cluster (Bower, Lucy \& Ellis
1992), whose common explanation is in terms of a sequence of passively evolving coeval objects of
decreasing metallicity. The number of ramp cycles (of 60s or 120s) and S/N reached at 6 $\mu$m are
shown in Table~1. The spectra were extracted within a fixed aperture (3.6"$\times$18" for SL and
10.2"$\times$10.4" for LL) and calibrated using our own software, tested versus the {\tt SMART}
software package (Higdon et al. 2004), as described in detail in Bressan et al. (2006a). For SL
observations, the sky background was removed by subtracting observations taken in different orders,
but at the same nod position. LL segments were sky-subtracted by differentiating the two nod
positions. Since the adopted IRS pipeline (version S12) is specifically designed for point source
flux extraction, we have devised a new procedure to flux calibrate the spectra, that exploits the
large degree of symmetry that characterizes the light distribution in early-type galaxies. The
intrinsic surface brightness profile has been derived by fitting to the data a bi-dimensional model
convolved with the point spread function (PSF) at several wavelengths. This procedure has a twofold
advantage because it allows us both to correctly reconstruct the intrinsic SED, and to recognize
whether a particular feature is spatially extended or not. Since for the LL segment the above
procedure is quite unstable, we have preferred to fix  one of the parameters of the profile
(usually R$_c$) to the value derived in the nearby wavelength region of the SL segment.
\begin{figure}
\centerline{\resizebox{0.9\textwidth}{!}{
\psfig{figure=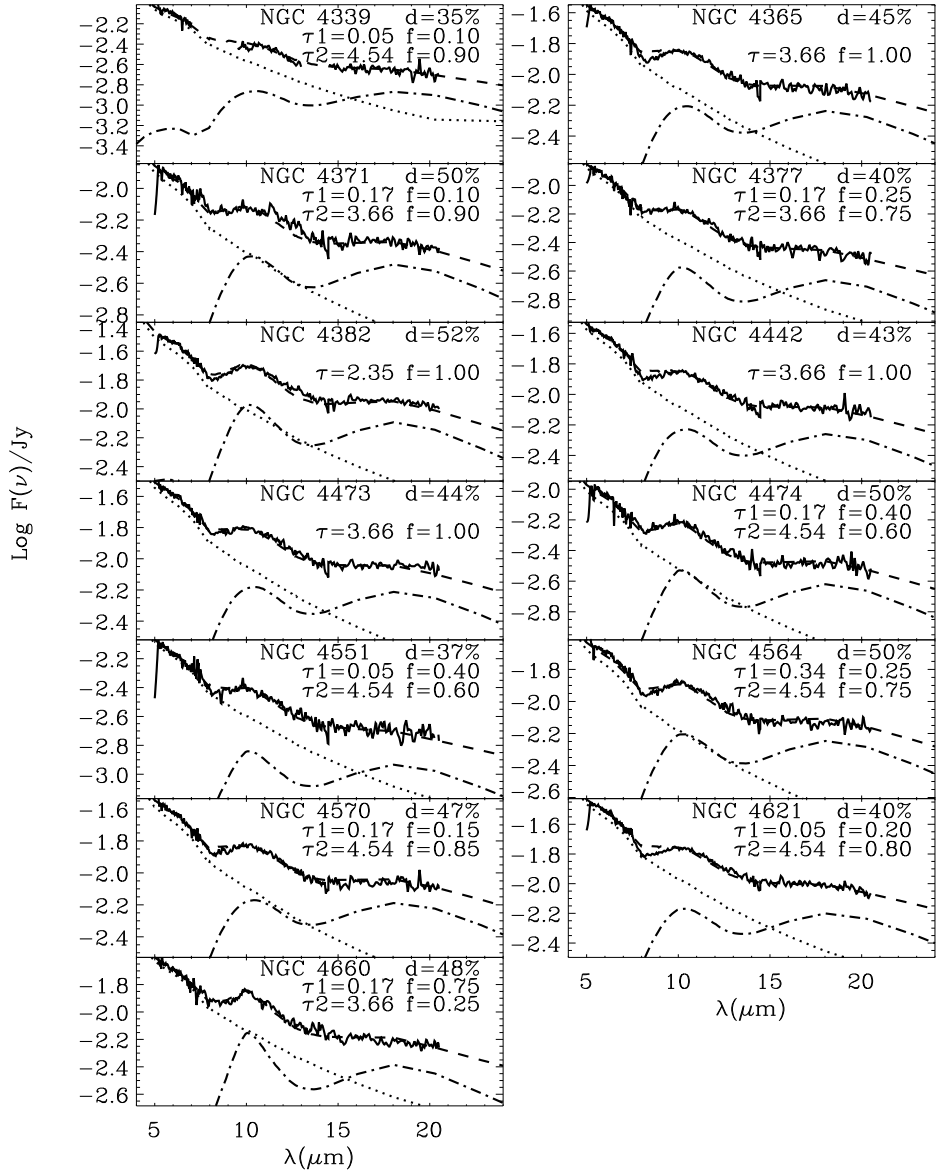} }}
\vskip 15pt
\caption{IRS spectra (solid lines) of
passively evolving early-type galaxies in the Virgo cluster.
Superimposed are best fits (dashed line) obtained by means of simple models
composed of an M giant star spectrum (dotted line) from
(MARCS, Gustafsson et al. 2002)
and a dusty silicate circumstellar envelope (dot-dashed line)
from Bressan et al. (1998).
Data and models are normalized at 5.5$\mu$m.
The fractional contribution of the circumstellar envelope at 10$\mu$m ("d")
and its optical depth at 1$\mu$m ("$\tau$") are indicated in the figure.
In those cases where a combination of  two dusty envelopes is needed,
their optical depth at 1$\mu$m and relative fractions ("f") are specified.
}
\label{passive}
\end{figure}

\section{Passive early-type galaxies.}

Thirteen galaxies (76\%) of our sample have a MIR spectrum that is dominated by the presence of a
broad emission feature above 10$\mu$m, {\sl without any other narrow emission feature}. The spectra
of these galaxies are collected in Figure \ref{passive}. The analysis of the IRS spectra indicates
that the  {\sl 10$\mu$m feature} has  an extended spatial distribution, consistent with that
obtained in the spectral range dominated by stellar photospheres (below 8$\mu$m). This has been
confirmed by the analysis of {\it Spitzer} IRS Peak-Up imaging observations in the blue (16$\mu$m)
filter of selected galaxies (Annibali et al. in preparation). It is also in agreement with previous
ISOCAM observations that indicated spatially resolved emission at both 6.7 and 15 $\mu$m (Athey et
al. 2002, Ferrari et al 2002, Xilouris et al. 2004). In view of these considerations and based on
preliminary fits with our models of passively evolving old simple stellar populations, we have
argued that we have detected the 10$\mu$m features, due to silicate emission from the circumstellar
envelopes of mass losing AGB stars, as predicted by Bressan et al. (1998).

However, the observed 10$\mu$m feature appears broader than that predicted by the models of Bressan
et al. (1998). A rather better fit is obtained by a simple superposition of an M giant model
spectrum (MARCS, Gustafsson et al. 2002), meant to represent the photospheric contribution of the
red giant population, and dust (silicate) emission from {\it moderately thick} circumstellar
envelope models (Bressan et al. 2006b), as shown in Figure \ref{passive}. It is surprising that the
observations indicate that the major contribution comes from envelopes in a quite narrow range of
optical depths ($\tau_{1{\mu}m}$ $\simeq 3-5$), at variance with isochrones that account for a
distribution of envelopes of varying optical depth. Possible ways out of this discrepancy are
either the inclusion of a different dust mixture and/or the use of a more detailed description for
the advanced phases (e.g. the recent AGB evolution by Girardi \& Marigo 2006).

As far as the dust mixture is concerned, there is evidence
that Al dust grains are present in circumstellar envelopes of AGB stars
(Blommaert et al. 2006, Lebzelter et al. 2006).
In the sample of  Bulge AGBs
observed by Blommaert et al. (2006) with ISO,
Al$_2$O$_3$ constitutes a large fraction of the
obscuring material (up to 100\% in many cases). Since Al$_2$O$_3$ has an emission  maximum at
about 12$\mu$m, one may suspect that a suitable combination of silicate and
Al$_2$O$_3$ dust grains may broaden the 10$\mu$m emission feature
to the level of reproducing
the observations of our passive galaxies.

A breakthrough in this respect is constituted by the {\it Spitzer} spectroscopy of AGB stars in the
globular cluster 47-Tuc (Lebzelter et al. 2006). This work highlights how the dust content changes
during the ascent of the AGB: at the beginning only a tiny MIR excess (at $\lambda\sim$12$\mu$m),
due to amorphous Al$_2$O$_3$ and crystalline MgAl$_2$O$_4$, is observed; then the excess due to
silicates begins to show up and finally dominates the MIR spectra (at $\lambda\sim$10$\mu$m) of the
most luminous and mass-losing AGB stars.

\begin{figure}
\centerline{\resizebox{0.7\textwidth}{!}{
\psfig{figure=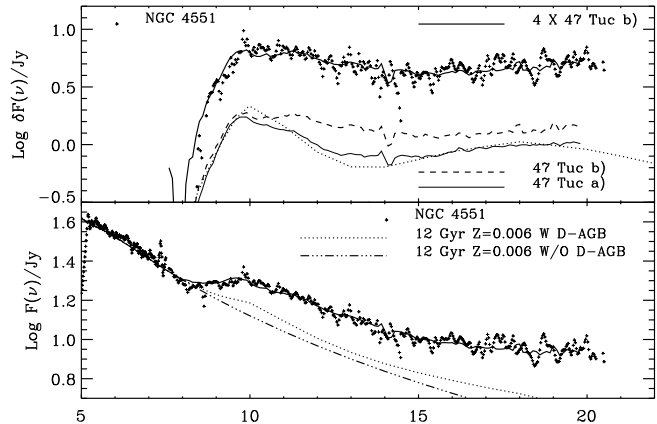,angle=90}}}
\caption{Comparison of IRS spectra of the early-type galaxy NGC 4551
with the semi-empirical one of 47-Tuc and SSP models with
and without dusty AGB envelopes.
The upper panel shows the MIR excess with respect to
the old SSP model computed without dusty AGB envelopes
while the lower panel show absolute values, normalized
at 5.5$\mu$m (see text for details).}
\label{excess}
\end{figure}

Besides that, these observations constitute a potential test for our SSP models. By summing up the
observed spectra we can obtain the {\it integrated} MIR light of AGB stars in 47-Tuc. We can then
subtract the {\it integrated} photospheric component to obtain the {\it integrated} MIR excess,
that can be directly compared with our observations and models. We have thus re-analyzed the
original data of Lebzelter et al. (2006), co-added the spectra and subtracted a pure photosphere
model normalized at 8$\mu$m (note that this is different from the excess derived by Lebzelter et
al., that instead have used a blackbody  fit to the NIR flux of the single stars). The excess is
shown in the upper panel of Figure \ref{excess}. The thin solid line is the observed excess
obtained by simply co-adding the spectra observed by Lebzelter et al. (2006). In the same panel the
dotted line is the MIR excess of a 12 Gyr old SSP for Z=0.006, as suggested from the abundance
pattern observed in 47-Tuc (Gratton et al. 2006). The SSP is normalized to the observed integrated
flux of 47-Tuc in the K band. The crosses are the measured excess of one of our sample galaxies,
NGC 4551, after normalization
of its spectrum to that of the SSP at 5.5$\mu$m.
However the sample observed by
Lebzelter et al. (2006) is only a subsample of the variable stars
detected by Lebzelter \& Wood (2005), lacking in particular the
majority of lower luminosity and shorter period variables.
To get a more representative
MIR spectrum of 47-Tuc, we have weighted
the spectra of Lebzelter et al. (2006),
with the number of neighboring variables in the
K-Log(Period) plane as derived from Lebzelter \& Wood (2005).
The resulting excess is the dashed line -labelled b)- in the
upper panel.
Finally, the solid thick line is the latter excess -b)-
multiplied by a factor of four, to match the observations of NGC 4451.

In the lower panel we show the {\it integrated} spectra normalized to the old SSP. The three
dot-dashed line represents the pure photospheric SSP spectrum used to estimate the MIR excess.

From this simple exercise we may draw a few interesting conclusions.
We expect that the real 47-Tuc integrated spectrum
is characterized by typical silicate emission, because it is
dominated by its brightest objects.
In fact already from the spectra of
Lebzelter et al. (2006) it appears that
the excess due to silicates is about one order of magnitude larger
than that due to Al grains. Nevertheless,
the contribution of the less luminous but more numerous
variables, mainly showing Al$_2$O$_3$ emission,
has the effect of broadening the 10$\mu$m peak.

In particular an excess four times larger than that obtained for 47 Tuc,
would fairly well match the shape of that
observed in the galaxy NGC 4551 (and other passive galaxies of Fig. \ref{passive}
as well, with slightly different factors).
This suggests that the mismatch between the shape
of the model spectra and the observations could be mainly caused
by the lack in the model of Al dust grain emission arising from
low mass loss rate AGB stars.

On the other hand, this experiment shows that the comparison of SSPs models with integrated globular 
cluster spectra must be done with care, because of the large stochastic effect introduced by the
very small number of bright AGBs. In fact we have checked that by only shifting a single bright AGB
star from its post-pulse luminosity dip to its quiescent interpulse luminosity, results in a
variation of about 30\% in the integrated excess at 10 $\mu$m.

Given the above results, we are now computing new isochrones and SSP models that account for a more
realistic description of the AGB phase and of their dusty envelopes. Particularly useful are the
new models by Girardi \& Marigo (2006) that are able to predict not only a more detailed evolution
in the HR diagram, but also the transition between the first overtone and the fundamental mode of
pulsation, that seems to separate the pure Al dust sequence from the silicate dominated dust
sequence (Lebzelter et al. 2006).

Finally we need to analyze the luminosity difference in the MIR excess between early-type galaxies
and 47-Tuc. This difference can be explained as due to the higher metallicity of the selected
early-type galaxies. In fact the MIR excess of the brightest 47-Tuc variable, V1, requires a very
thin circumstellar envelope, $\tau_{0.55{\mu}m}\simeq$0.15 (van Loon et al. 2006), corresponding to
$\tau_{1{\mu}m}\simeq$0.04 in the spectral region of maximum photospheric emission. It is easy to
show that as long as $\tau_{1{\mu}m}\ll$1 the MIR excess over an M-giant spectrum increases linearly
with $\tau_{1{\mu}m}$. Narrow band indices, colors and a recent {\it direct measurement} based on
our Spitzer data for NGC 4435 (Panuzzo et al. 2007) suggest that the metallicity of luminous ETGs
is solar or beyond. Thus the four times larger MIR excess found in early-type galaxies can be
entirely due to the larger dust/gas ratio of old AGB stars with a metallicity which is around four
times that of 47-Tuc. However the metallicity may affect other physical processes that are
relevant to the prediction of the dusty AGB phase, such as the pulsation properties, the mass-loss
rate and, for the younger populations, the set up of the third dredge-up phase.

In order to analyze the importance of all those processes and possibly to cope with the stochastic
effect, a much larger AGB sample must be considered. In particular Bulge stars constitute the most
natural sample to be analyzed and compared with stellar evolutionary models, because they are old
and their metallicity is approximately solar. Since a large region of the Bulge has already been observed
with Spitzer (IRAC and MIPS imaging) we are expecting new exciting results to come soon. We suggest
however that the best strategy to highlight (even modest) dust effects would be to complement IRAC
and MIPS observations with IRS Blue Peak-Up observations, since the latter band
($\lambda_c\sim16\mu$m) is centered right in the middle  of the MIR excess of mass-losing stars.
\begin{figure}
\centerline{\resizebox{0.9\textwidth}{!}{
\psfig{figure=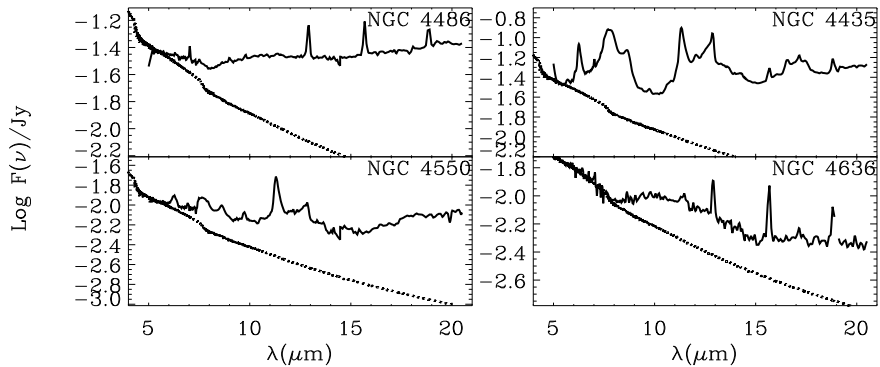,clip=}
}}
\caption{IRS spectra (solid lines) of
active early-type galaxies in the Virgo cluster.
In this case, only the normalized M giant star spectrum
is shown (dotted line).}
\label{active}
\end{figure}

\section{Active galaxies.}

Among bright Virgo cluster ETGs observed by our team, a significant number are not passively
evolving but show various levels of activity. This is clearly the case for the other four
galaxies shown in Figure \ref{active}. They constitute 24\% of our sample. Spectra of NGC 4636 and
NGC 4486(M87) are dominated by emission lines ([ArII]7$\mu$m, [NeII]12.8$\mu$m, [NeIII]15.5$\mu$m
and  [SIII]18.7$\mu$m), while those of  NGC 4550 and NGC 4435 by PAH emission (at 6.2, 7.7, 8.6,
11.3, 11.9, 12.7  and 16.4$\mu$m). The broad continuum features at 10$\mu$m (and perhaps at
18$\mu$m) in M87 are unresolved and their likely cause is silicate emission from the dusty torus
(Siebenmorgen et al. 2005, Hao et al. 2005). NGC 4435 shows also emission lines (and H2 S(3)
9.66$\mu$m, H2S(2) 12.3$\mu$m and H2S(1)17.04$\mu$m). 

The analysis of the MIR spectrum of NGC 4435 together with its global SED,
indicates that the nucleus of this galaxy is in a post-starburst phase.
The {\it rejuvenation episode} is of low intensity
($\sim$0.3\% of the total mass of the galaxy) 
and was likely triggered by the interaction with NGC 4438
(Panuzzo et al. 2007).
The strengths of MIR nebular emission lines 
imply that $\sim$98\% of the ionizing flux 
is of stellar origin.
At the same time they
allow an accurate determination of the 
metallicity of the gas in the nuclear disk which turns out to be of solar composition.

\section{Conclusions}
We have obtained Spitzer mid infrared IRS spectra of early-type galaxies selected along the
colour-magnitude relation of the Virgo cluster. In most of the galaxies (76\%) the emission looks
spatially extended and presents an excess longward of 10$\mu$m, which is likely due to silicate
emission from mass losing evolved stars. The observed MIR excess is about four times larger than
that estimated for 47-Tuc, on the basis of IRS spectroscopy of its brightest AGB stars. This could
be the consequence of a higher dust/gas ratio, in agreement with the higher metallicity of
early-type galaxies. However, since other important physical processes are affected by metallicity,
a thorough comparison between new Spitzer data and new models of the advanced phases are needed.
Ideal workbenches for the models would be colour-magnitude and colour-colour diagrams of the Bulge
stellar population, especially in the wavelength range  of the IRS Blue Peak-Up.

In the remaining fraction of galaxies (24\%) we detect signatures of activity
at various levels. The analysis of the IRS spectrum of NGC 4435
testifies to the superb capability of 
Spitzer to probe the nature of this type of activity.

A detailed comparison of these results
with those obtained for field early type galaxies will certainly cast light on the
role of environment in the galaxy evolution process.

\acknowledgements 
This work is based on observations made with the Spitzer Space Telescope, which
is operated by the JPL, Caltech
under a contract with NASA.
A. B., G.L. G. and L. S.  thank INAOE for warm hospitality.



\begin{thebibliography}{}
\bibitem[Athey et al.(2002)]{2002ApJ...571..272A} Athey, A., Bregman, J., Bregman, J., Temi, P., \& Sauvage, M.\ 2002, \apj, 571, 272
\bibitem[1992]{Bowe92} Bower, R. G., Lucey, J. R., Ellis, R. S. 1992, MNRAS, 254, 601
\bibitem[Blommaert et al.(2006)]{2006A&A...460..555B} Blommaert,
J.~A.~D.~L., et al.\ 2006, \aap, 460, 555
\bibitem[Bressan et al.(2006)]{2006astro.ph..4068B} Bressan, A., et al.\
2006, ArXiv Astrophysics e-prints, arXiv:astro-ph/0604068
\bibitem[Bressan et al.(2006)]{2006ApJ...639L..55B} Bressan, A., et al.\
2006, \apjl, 639, L55
\bibitem[Bressan et al.(2001)]{2001ApSSS.277..251B} Bressan, A., Aussel, H., Granato, G.L. et al.\ 2001, \apss, 277, 251
\bibitem[1998]{Bres98}  Bressan, A., Granato, G.L., Silva, L. 1998, AA, 332, 135
\bibitem[Ferrari et al.(2002)]{2002A&A...389..355F} Ferrari, F., Pastoriza, M.~G., Macchetto, F.~D. et al. \ 2002, \aap, 389, 355
\bibitem[Girardi \& Marigo(2006)]{2006astro.ph..9626G} Girardi, L., \&
Marigo, P.\ 2006, ArXiv Astrophysics e-prints, arXiv:astro-ph/0609626
\bibitem[Gratton et al.(2004)]{2004ARA&A..42..385G} Gratton, R., Sneden,
C., \& Carretta, E.\ 2004, \araa, 42, 385
\bibitem[Gustafsson et al.(2002)]{Conf} Gustafsson B. et al. 2002, ASP Conf. Ser. Vol. 288, p. 331
\bibitem[Hao et al.(2005)]{2005ApJ...625L..75H} Hao, L., et al.\ 2005, \apjl, 625, L75
\bibitem[2004]{Higd04} Higdon, S.J.U. et al. 2004, PASP, 116, 975
\bibitem[Ho et al.(1997)]{hofi97} Ho, L. C., Filippenko, A. V., \& Sargent, W. L. 1997, ApJS, 112, 315
\bibitem[2004]{Houc04} Houck, J.R. 2004, ApJS, 154, 18
\bibitem[Kaneda et al.(2005)]{2005ApJ...632L..83K} Kaneda, H., Onaka, T.,  \& Sakon, I.\ 2005, \apjl, 632, L83
\bibitem[Lebzelter et al.(2006)]{2006ApJ...653L.145L} Lebzelter, T., Posch,
T., Hinkle, K., Wood, P.~R., \& Bouwman, J.\ 2006, \apjl, 653, L145
\bibitem[Lebzelter \& Wood(2005)]{2005A&A...441.1117L} Lebzelter, T., \&
Wood, P.~R.\ 2005, \aap, 441, 1117
\bibitem[Panuzzo et al.(2007)]{2006astro.ph.10316P} Panuzzo, P., et al.\
2007, ApJ in press (astro-ph/0610316)
\bibitem[Piovan et al.(2003)]{2003A&A...408..559P} Piovan, L., Tantalo, R.,
\& Chiosi, C.\ 2003, \aap, 408, 559
\bibitem[Siebenmorgen et al.(2005)]{2005A&A...436L...5S} Siebenmorgen, R.,
Haas, M., Kr{\"u}gel, E., \& Schulz, B.\ 2005, \aap, 436, L5
\bibitem[Xilouris et al.(2004)]{2004A&A...416...41X} Xilouris, E.~M. et al. \ 2004, \aap, 416,  41
\end{thebibliography}
\end{document}